\documentclass{aastex}
\usepackage{spr-astr-addons}
\usepackage{url}
\usepackage{graphicx,natbib}
\usepackage{amssymb}

\def\iue{\it IUE\/}
\def\units{$10^{-11}\mathrm{erg}\hspace{0.8mm}\mathrm{s}^{-1}\mathrm{cm}^{-2}\mathrm{\AA}^{-1}$}

\begin{document}

%\onecolumn

\title{Spectrophotometric variability of the magnetic CP star\\
       ~56~Arietis in spectral region from 1950 to 3200 \AA}

\author{Nikolay A. Sokolov}
\affil{Central Astronomical Observatory at Pulkovo,
St. Petersburg 196140, Russia}

\email{n\_sokolov58@mail.ru}

%\thanks{Based on $IUE$ Newly Extracted Spectra (INES) data from
%the $IUE$ satellite.}

\begin{abstract}
The spectrophotometric variability of the magnetic CP star
56~Arietis (56~Ari) in the ultraviolet spectral region from 1950 to
3200~\AA\ is investigated. This study is based on the archival {\it
International Ultraviolet Explorer\/} data obtained at different
phases of the rotational cycle. The brightness of 56~Ari is not
constant in the investigated wavelengths over the whole rotational
period. The monochromatic light curves continuously change their
shape with wavelength. This indicates that we do not observe a truly
'null wavelength region' where the monochromatic light curve has a zero amplitude.
Probably, an uneven surface distribution of  silicon and iron mainly influences
the flux redistribution from the far-UV to near-UV spectral regions, although
additional sources of opacity may be involved. The redistribution of the flux at phase 0.25 is connected with the nonuniform distribution of silicon on the stellar surface
of 56~Ari. On the other hand, the redistribution of the flux at phase 0.65 is quite complex, because there are additional blocking and redistribution of the flux by iron lines in the near-UV spectral region.
\end{abstract}


\maketitle
%\onecolumn

%%%%%%%%%%%%%%%%%%%%%%%%%%%%%%%%%%%%%%%%%%%%
%% MAINMATTER
%%%%%%%%%%%%%%%%%%%%%%%%%%%%%%%%%%%%%%%%%%%%

\section{Introduction}

The magnetic Chemically Peculiar (mCP) star 56~Ari (SX Ari,
HD~19832, HR~954) belongs to the Si-group. The star, classified as B8~Si by
\citet{Renson1991}, is a periodic spectrum \citep{Deutsch1947}, photometric
\citep{Provin1953} and magnetic \citep{Borra1980} variable, with a relatively
short period of 0.728 days. The light variability can be generally expected to be
explained by the inhomogeneous abundance of several chemical elements which is
observed in the atmosphere of 56~Ari. \citet{Jamar1978} had investigated the
ultraviolet spectral variations of six CP Si stars, including 56~Ari, by means of
observations obtained from the Sky Survey Telescope (S2/68) on the TD1 satellite.
In the case of 56~Ari, the author established that the light variability in
the short wavelength region is clearly in phase opposition with
the remaining part of the spectrum with a `null wavelength region` at 1600~\AA.
But this conclusion is naturally only valid in wide band photometry because
the spectra intercrosses many times at other wavelengths
\citep[see Fig~1 of][]{Jamar1978}.
The author has noted that it is impossible to say if this effect is real or
only the result of a poor signal-to-noise ratio.
\citet{Stepien_Czechowski} have analyzed the spectrophotometric behavior of
this star, using the {\it International Ultraviolet Explorer\/} ({\iue}) data.
Unfortunately, the authors restricted the investigation of the
long-wavelength spectra from the {\iue} data only in the three
spectral regions at $\lambda\lambda$~2100, 2500 and 3000~\AA.
Besides, the authors have called the light curves
"monochromatic", although they were determined by averaging the fluxes
over intervals [$\lambda$-10,~$\lambda$+10]~\AA\ for given $\lambda$.

Recently, \citet[ hereinafter Paper I]{Sokolov2006} analyzed
the spectrophotometric variability of 56~Ari in the far-ultraviolet (far-UV)
spectral region from 1150 to 1980~\AA. The author established that
the brightness of 56~Ari is not constant in the investigated wavelengths
over the whole rotational period, although, the double-wave light variations in
the far-UV spectral region are in antiphase to the visual spectral region.
Moreover, the author involved the additional sources of
variable opacity in order to explain the existence of the second
minimum at phase 0.65 of the monochromatic light curves.
He concluded that the inhomogeneous surface distribution of iron seems to have
influence on the light variations in the far-UV spectral region.
Perhaps the most interesting and controversial aspect of 56~Ari is its
spectrophotometric variability in the near ultraviolet (near-UV) spectral
region from 1950 to 3200~\AA. In this spectral region, the photometric level
is strongly influenced by the multiplets of iron lines. These multiplets are
presented in the spectra of CP~stars \citep[and references therein]{Jamar1978}.

In this paper, the low-dispersion spectra of mCP star 56~Ari are
analyzed in detail using long-wavelength data from the final {\iue}
archive. Moreover, the variability of selected features and spectral
lines in this spectral region can be established.

\section[] {Observational data}

\subsection[] {{\iue} spectra}

Three series of observations of 56~Ari obtained with Long Wavelength
Redundant (LWR) and Long Wavelength Prime (LWP) cameras were
retrieved from the final {\iue} archive.
\begin{enumerate}
\item The first one contains 18 LWR spectra obtained in December 1981.
\item The second one contains 8 LWP spectra obtained in February 1984.
\item The third one contains 7 LWP spectra obtained in August and
September 1990.
\end{enumerate}
In all cases, the spectra were obtained in the low-dispersion mode
through the large aperture (9.5$\arcsec$$\times$22$\arcsec$).
The spectra were extracted from {\iue} {\it Newly Extracted Spectra}
(INES) {\it Catalog} in which the flux in absolute units (cgs) is
given every 2.67~\AA\ for LWR and LWP cameras. The INES data from
{\iue} satellite are available from the INES Principal Center
website {\rm http://sdc.laeff.inta.es/ines/}.

The INES catalog contains a few spectra of 56~Ari obtained with the
small aperture. Unfortunately, there are systematic differences in
the LWR and LWP spectra with large and small apertures, because only
point-source large aperture spectra were used for the absolute
flux calibration \citep{Gonz2001}. In order to exclude these
systematic differences, only {\iue} spectra obtained with a large
aperture were used in our study. Moreover, the signal-to-noise ratio
of the last series observations is very poor, because the exposure
time is very small, and these spectra appear unsuitable for our
purpose. Additionally, the spectrum LWP~02833 was excluded, because
the exposure time is only 4.383 sec. Finally, we analyze 18 LWR and
7 LWP spectra, distributed quite smoothly over the rotational period.
In Table~1, each spectrum is presented by its number, exposure
time, Julian date of the observations and its corresponding phase obtained
using equation~(\ref{ephemeris}).

\begin{table} [t]
\small
\vspace{-2mm}
\centering
\begin{minipage}{84mm}
\caption{List of the spectral {\it IUE} observations of 56~Ari obtained
 in the low-dispersion mode through the large aperture.}
\begin{tabular} {lccc}
\tableline
\noalign{\smallskip}
$IUE$~images&Exp. Time&Julian date&Phase\\&(sec.)&2,440,000+&\\
%\noalign{\smallskip}
\tableline
\noalign{\smallskip}
LWR~12194$^*$ & 19.190 & 4962.2839 & 0.320\\
LWR~12195$^*$ & 19.190 & 4962.3300 & 0.384\\
LWR~12196$^*$ & 19.190 & 4962.3754 & 0.446\\
LWR~12197$^*$ & 19.190 & 4962.4212 & 0.509\\
LWR~12198$^*$ & 19.190 & 4962.4668 & 0.572\\
LWR~12199$^*$ & 19.190 & 4962.5124 & 0.634\\
LWR~12200$^*$ & 19.190 & 4962.5589 & 0.698\\
LWR~12201$^*$ & 19.190 & 4962.6129 & 0.772\\
LWR~12202$^*$ & 19.190 & 4962.6566 & 0.832\\
LWR~12203$^*$ & 19.190 & 4962.7073 & 0.902\\
LWR~12204$^*$ & 19.190 & 4962.7505 & 0.961\\
LWR~12205$^*$ & 19.190 & 4962.7973 & 0.026\\
LWR~12206$^*$ & 19.190 & 4962.8418 & 0.087\\
LWR~12207$^*$ & 19.190 & 4962.8859 & 0.147\\
LWR~12213$^*$ & 19.190 & 4963.6250 & 0.163\\
LWR~12214$^*$ & 19.190 & 4963.6757 & 0.232\\
LWR~12215$^*$ & 19.190 & 4963.7199 & 0.293\\
LWR~12216$^*$ & 19.190 & 4963.7657 & 0.356\\
%\noalign{\smallskip}
%\tableline
%\end{tabular}
%\end{minipage}
%\end{table}
%\setcounter{table}{0}
%\begin{table} [t]
%\small
%\centering
%\vspace{-2.0mm}
%\begin{minipage}{84mm}
%\caption{continued}
%\begin{tabular} {cccc}
%\tableline
%$IUE$~images&Exp. Time&Julian date&Phase\\&(sec.)&2,440,000+&\\
%\tableline
LWP~02754$^*$ & 18.586 & 5740.4630 & 0.392\\
LWP~02755$^*$ & 18.586 & 5740.5441 & 0.503\\
LWP~02756$^*$ & 18.586 & 5740.6188 & 0.606\\
LWP~02828$^*$ & 18.586 & 5752.4030 & 0.795\\
LWP~02829$^*$ & 18.586 & 5752.5171 & 0.952\\
LWP~02831$^*$ & 18.586 & 5753.3655 & 0.117\\
LWP~02832$^*$ & 18.586 & 5753.4398 & 0.219\\
LWP~02833 &  4.383 & 5753.5257 & 0.337\\
LWP~18613 &  4.383 & 8122.5789 & 0.970\\
LWP~18614 &  3.973 & 8122.6225 & 0.029\\
LWP~18615 &  2.744 & 8122.7426 & 0.194\\
LWP~18616 &  3.563 & 8122.8206 & 0.301\\
LWP~18617 &  3.563 & 8122.8714 & 0.371\\
LWP~18847 &  3.563 & 8156.7485 & 0.912\\
LWP~18849 &  3.563 & 8156.9708 & 0.217\\
\tableline
\end{tabular}
\end{minipage}
\tablenotetext{*}{Spectra used in our analysis.}
\end{table}

\begin{figure*}
\rotate
\vspace{1.5mm}
\centerline{\includegraphics[width=140mm]{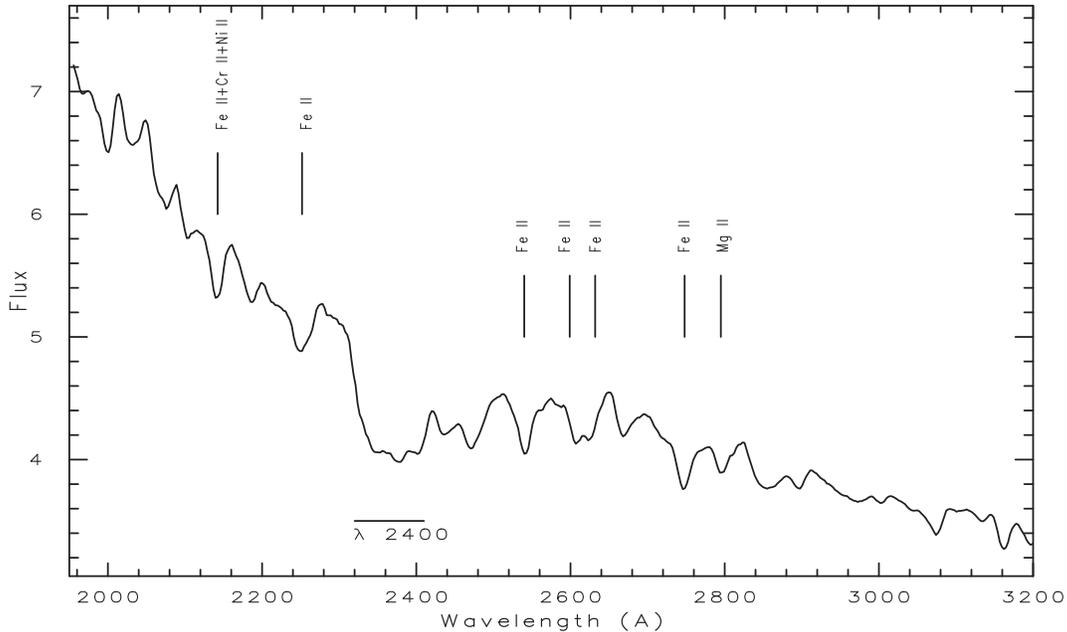}}
\caption{The average energy distribution in {\units} for 56~Ari.
 The prominent spectral lines and features are shown by {\it vertical} and
 {\it horizontal lines}, respectively.
\label{mean}}
\end{figure*}

\subsection[] {The period variations of 56~Ari}

Some mCP stars, including 56~Ari, displayed an increase of their rotational
periods \citep{Mikulasek2008}. The rotational period of 56~Ari has been
studied by various investigators over 48 yr \cite[see e.g.][]{Adelman2001}.
\citet{Musielok1988} had analyzed $UBV$ and $uvby\beta$ photometry for
the interval JD~2434322-2447176, which covers over 17~660 rotational cycles.
He found an increase in the rotational period
of 4~s per 100 yr from an analysis of the $U$ photometric data.
Possibly, for this star the period's increase is also accompanied by changes of
its $UBV$ light curves \citep{Adelman_Fried, Ziznovsky2000}. The authors expected
the observed period changes because of free body precession \citep{Shore1976}.
\citet{Adelman2001} had studied all available photometric data for this star from
1952 to 2000. They confirmed the increase in the rotational period, but with
a rate of about 2~s per 100 yr.
The authors pointed out that there is evidence for a second
period of about 5 yr attributed to the precession of the axis of rotation.

In Paper~I the phases of the monochromatic light curves in the far-UV
were computed by using the ephemeris with constant period obtained by \\
\citet{Adelman_Fried} and linearly changing period obtained by \citet{Adelman2001}.
In most cases, the monochromatic light curves in the far-UV exhibit the maximum
of the flux at phase 0.0 with the ephemeris of \citet{Adelman_Fried}:
\begin{equation}
{\mathrm JD(U,B~min)}=2439797.586+{0^{\mathrm d}}\hspace{-0.9mm}.727902 E.
\label{ephemeris}
\end{equation}
With this ephemeris, the monochromatic light curves of 56~Ari show,
as a rule, two maxima and two minima in the far-UV spectral region
(see Fig.~1 of Paper~I). Based on this fact, in our investigation the phases
were computed using equation~(\ref{ephemeris}).

\section[] {Data analysis}

As well as for short-wavelength {\iue} monochromatic light curves,
we used the linearized least squares method.
An attempt was made to describe the light curves in
quantitative way by adjusting a Fourier series.
This method is described by \citet{North1987} and assumes that
the shape of the curve has the form:
\begin{equation}
F(\lambda, T^{'})=A_{0}(\lambda) + \sum_{i=1}^{n}
A_{i}(\lambda)\cos(\omega~i~T^{'} +\phi_{i}(\lambda))
\label{fourier}
\end{equation}
where $F(\lambda, T^{'})$ is a flux for the given $\lambda$, $T^{'}$=~$T$~-~$T_{0}$
and $\omega$~=~2$\pi$/$P$.
The $T_{0}$ and $P$ are zero epoch and rotational period of
the ephemeris, respectively and $n$ is the total number of harmonics.
The coefficients $A_{0}(\lambda)$ of the fitted curves are defining
the average energy distribution over the cycle of the variability.
From several scans distributed over the rotational period, it is possible to
compute the light curves at different wavelengths.
In all wavelengths the data could be fitted by Fourier series limited
to $n$=3, that is, by the fundamental frequency and its first two harmonics.
This procedure can be partially accounted by considering that within the
accuracy of the measurements a cosine wave and its first two harmonics appear
to be generally adequate to describe the monochromatic light curves in
the far-UV (see Paper~I).
Additionally, the least squares fit with three-frequency cosine function was applied
to the $UBV$ light curves obtained by \citet{Fried2003}. The computations showed
that within the accuracy of the measurements it is also true for the light
variations in the visible region.

The least squares fit was applied to separate long-wavelength {\iue}
monochromatic light curves. An error analysis showed that the errors
in the coefficients $A_{0}(\lambda)$ and $A_{i}(\lambda)$ of the fitted curves
are not more 0.05 and 0.07 in the units {\units}, respectively.
Although, the standard deviations of the residual scatter around the fitted curves ($\sigma_{res}(\lambda)$) varies from 0.03 to 0.2~$\times$~{\units}
in the investigated wavelengths.
The maximal errors of the coefficients $A_{0}(\lambda)$ and $A_{i}(\lambda)$
as well as in $\sigma_{res}(\lambda)$ are in the blue and red parts of {\iue}
spectrum. Probably, it is connected with uncertainties of the fluxes in both
ends of the spectrum as presented in INES database. Thus we limited our
investigation to the wavelength region between 1950 to 3200~\AA.
In order to minimize the uncertainties in the coefficients of the fitted curves
the fluxes were determined by averaging three nearest fluxes for a given $\lambda$:
\begin{equation}
 F(\lambda) = \frac{F(\lambda-\lambda_{step}) + F(\lambda) + F(\lambda+\lambda_{step})} {3},
\label{Mean_Flux}
\end{equation}
where $\lambda_{step}$ is equal 2.669~\AA\ (see Sect.~2).

As far as the errors in $F(\lambda)$ are concerned, we computed them by taking
into account the the errors in the fluxes as presented in INES $Catalog$,
according to the standard propagation theory of errors.

\begin{figure}
\vspace{-15.0mm}
\centerline{\includegraphics[width=80mm]{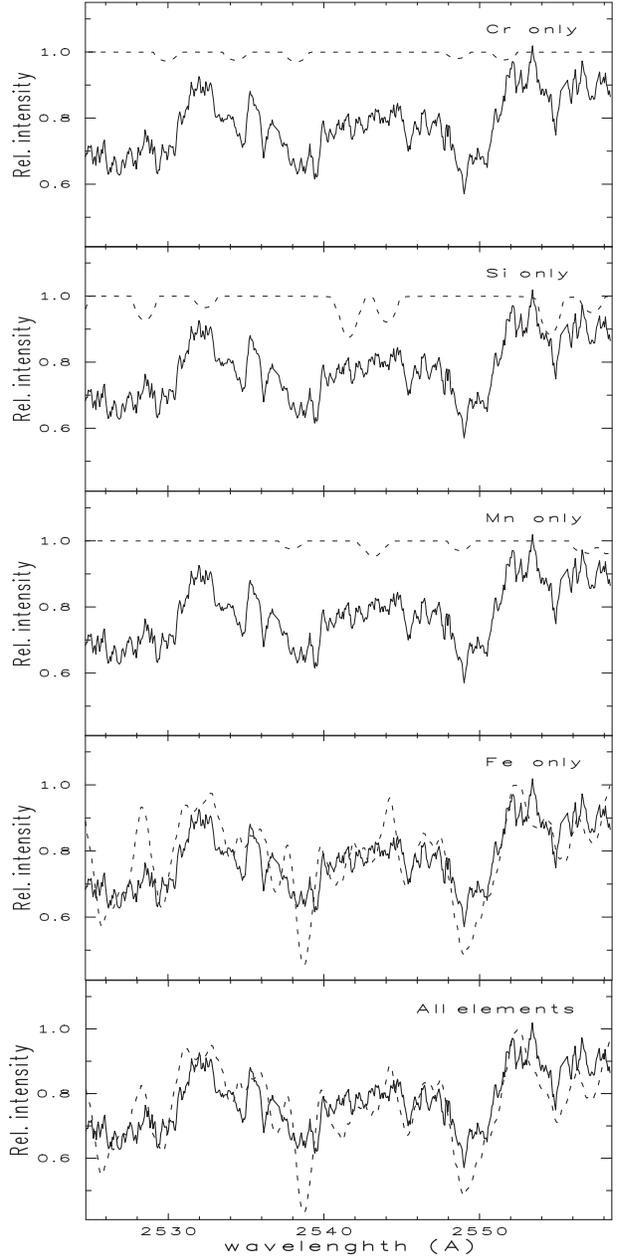}}
\caption{{\iue} high-dispersion spectrum of 56~Ari ({\it solid line}) compared to
 synthetic spectra ({\it dashed line}). The synthetic spectrum labelled 'All elements'
 was calculated with all elements. The 'Cr only' is a synthetic spectrum showing
 only the Cr lines. Similar single-element spectra were calculated for Cr, Si,
 Co, Mn, Ni and Fe (see the text).
\label{synt_spec}}
\end{figure}

\begin{figure}
\vspace{-15.0mm}
\centerline{\includegraphics[width=80mm]{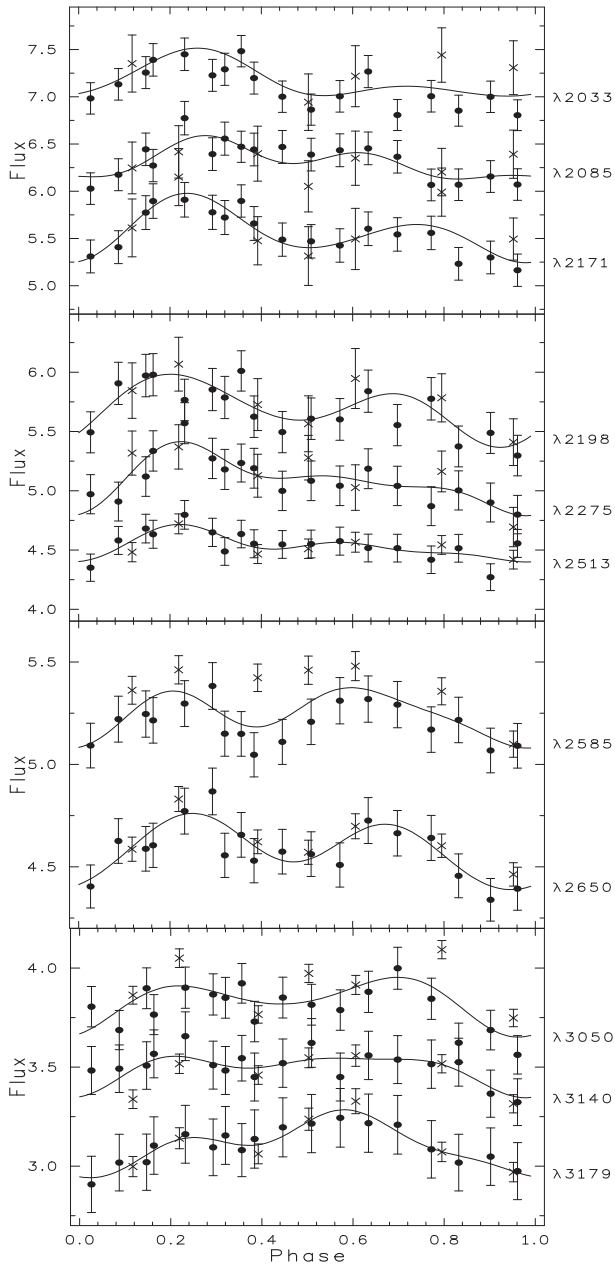}}
\caption{Phase diagrams of the monochromatic light curves in {\units} for 56~Ari.
 In the all plots, {\it closed circles} and {\it crosses} represent the data
 from the first and second series, respectively.
 To avoid overlapping the vertical shift on the constant value
 of some curves was used (see text).
 The {\it solid lines} are the least square fits.
\label{mono_light}}
\end{figure}

\subsection[] {Identification of the observed depressions and features in
the spectrum of 56~Ari}

Figure~\ref{mean} displays the average energy distribution ($A_{0}(\lambda)$)
of 56~Ari over the cycle of the variability in the spectral region from 1950 to
3200~\AA. First of all, we need to identify which elements are
responsible for depressions centered at $\lambda\lambda$~2140, 2250,
2540, 2607, 2624, 2747 and 2800~\AA.
In order to identify which element is responsible for these depressions
the synthetic single-element spectra were calculated for Si, Fe, Ni, Cr, Co, Mg
and Mn with \citet{Piskunov1992} program {\sc synth}. Typically, the lines
of these elements dominate at the selected spectral regions.
Also, the synthetic spectra with all elements included were computed.
The information about spectral lines were taken from the Vienna
Atomic Line Database \citep[VALD-2,][]{Kupka1999}. The atmospheric
parameters were estimated as a average from available sources.
\citet{Sokolov1998} estimated the value of $T_{\rm eff}$~=~12460~K from the
slope of the energy distribution in the Balmer continuum near the
Balmer jump. \citet{Adelman_Rayle} determined the values of $T_{\rm eff}$~=~12850~K, log~$g$~=~4.0, and $v_{\rm micro}$~=~2~km~${\rm s}^{\rm -1}$ from the fit of
the energy distribution in visual spectral region with the prediction of
the metal-rich models.
Recently, new determinations of $T_{\rm eff}$ of 23 mCP stars were
obtained by \citet{Lipski_Stepien} from a fit of metal enhanced model atmospheres
to the observed spectral energy distribution from UV to red. In the case of
56~Ari, they determined the value of $T_{\rm eff}$~=~12250~K. On the other hand,
the temperatures estimated separately at UV and visual spectral regions are equal 12250~K and 12750~K, respectively \citep{Lipski_Stepien}.
The temperature of 56~Ari found via photometry is equal to 12600~K \citep{Adelman_Rayle}. The model computation was performed with $T_{\rm eff}$~=~12500~K, log~$g$~=~4.0 and $v_{\rm micro}$~=~2~km~${\rm s}^{\rm -1}$.
It should be noted that the program {\sc synth} used the grids of ATLAS9 model atmospheres (http://kurucz.harvard.edu/grids.html).
But, the VALD-2 allows one to select the spectral lines for models with abundances significantly different from the solar or scaled solar composition.
The best agreement between {\iue} high resolution spectrum of 56~Ari at
phase 0.027 (SWP~39680) in spectral region of the depression at
$\lambda$1775~\AA\ and the synthetic spectrum with all elements is
reached if the elements have solar scaled composition [M/H]=0 except
for Al and Si, their abundances were reduced to log~$N$/$N_{\rm total}$
of -7.0 and -3.5, respectively (see Paper~I). Note that the chemical composition
is not a very peculiar one, because the energy distribution at phase 0.0 is close
to the normal energy distribution of 56~Ari. Our computations for depressions
centered at $\lambda\lambda$~2140, 2250, 2540, 2607, 2624, 2747 and 2800~\AA\ were performed with same peculiar chemical composition. Additionally, the synthetic spectra
were broadened according to a projected rotational velocity of 96.0~km~${\rm s}^{\rm -1}$ \citep{Royer2002} with program ROTATE written for PC by \citet{Piskunov1992}.

The computed synthetic spectra showed that Fe(II) appears as the main
absorber at $\lambda\lambda$~2250, 2540, 2607, 2624 and 2747~\AA.
For example, the comparison of the {\iue} high-dispersion spectrum
of 56~Ari at phase 0.072 (LWP~18848) with the synthetic spectrum including all
elements as well as those including lines from only one element shows that the iron
is responsible for depression centered at $\lambda$~2540~\AA, as illustrated by Fig.~\ref{synt_spec}. Some discrepancy between the synthetic spectrum and
the observed spectrum can be explained by low values of a signal-to-noise ratio
for the spectrum obtained in the high-dispersion mode.
Although, a wrong continuum assumed can also be caused by this discrepancy.
The same comparison of the {\iue} high-dispersion spectrum with the synthetic spectra showed that mainly Fe, Cr and Ni are responsible for depression at $\lambda$~2140~\AA.
It is well known, that Mg~{\sc ii} resonance lines at $\lambda\lambda$~2795, 2798
and 2803~\AA\ are responsible for depression at $\lambda$~2800~\AA.
Our model calculation fully supports this result. Prominent depressions
of the flux are indicated on Fig.~\ref{mean} with their identification.

The large feature at $\lambda$~2350--2400~\AA\, which are strongly
enhanced in the spectrum of CP stars, are well seen in spectrum of
56~Ari. The lines of iron peak elements have a particularly important contribution
to opacity at $\lambda$~2350--2400~\AA. \citet{Adelman1993} had given strong
arguments supporting the idea that a large number the lines of
iron can explain the feature at $\lambda$~2350--2400~\AA.
Many CP~stars have here a very pronounced depression compared to
normal stars \citep{Stepien_Czechowski}.

\begin{figure} [t]
\vspace{1.5mm}
\centerline{\includegraphics[width=75mm]{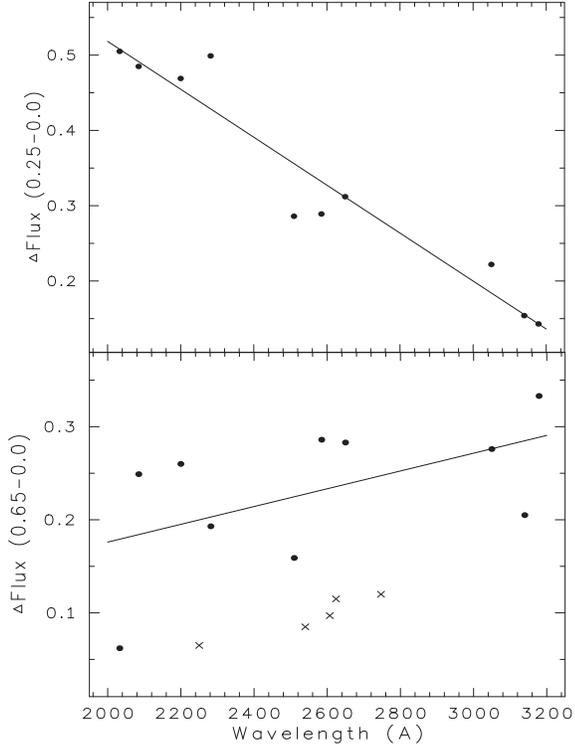}}
\caption{The differences of the fluxes of 56~Ari in
 {\units} with wavelength for three phases. The {\it top panel} shows the
 difference of the fluxes in the 'pseudo-continuum' at phases 0.25
 and 0.0; the {\it bottom panel} shows the difference of the fluxes in the
 'pseudo-continuum' ({\it circles}) and in the cores of Fe(II) lines
 ({\it crosses}) at phases 0.65 and 0.0. The {\it solid lines} are the linear
 least square fits.
\label{differ_flux}}
\end{figure}

\subsection[] {The monochromatic light variations in the pseudo-continuum}

First, it is necessary to fix the continuum in the low dispersion
{\iue} data. This is very difficult in the near-UV due to the lines
crowding. Nevertheless, one can find some high flux points located
at the same wavelengths in several spectra of 56~Ari. It should be
noted that such choice of the continuum might be a 'pseudo-continuum'.
However, there is no chance to reach the true continuum, if it occurs
at high flux points.
From several scans distributed over the rotational period one can produce
the light curves in different wavelengths. The light curves discussed below
will be called "monochromatic", although they were determined by
averaging three nearest fluxes for a given $\lambda$.
%As far as the errors on the fluxes are concerned, we computed them by taking
%into account the errors of separate fluxes from INES database,
%according to the standard propagation theory of errors.
%(see Eq.~\ref{Errors_Flux}).
Several monochromatic light curves in the 'pseudo-continuum' at different
wavelengths were formed. The examples of light curves together with the fitted
three-frequency cosine curves are shown in Fig.~\ref{mono_light}.
It should be noted that the vertical scales differ for each part of the figure.
In order to exclude overlapping of some curves the vertical shift on the constant
value was used. In this way, the curves at $\lambda\lambda$~2275 and 3179~\AA\
were shifted down to the values of -0.25 and -0.4~$\times$~{\units},
respectively. On the other hand, the curves at
$\lambda\lambda$~2033, 2198, 2585 and 3050~\AA\ were shifted up to the
values of +0.75, +0.25, +0.75 and +0.25~$\times$~{\units}, respectively.
Unfortunately, there are the systematical disagreements between LWR and LWP
cameras for some spectra and for some wavelengths (e.g. at $\lambda$~2585~\AA),
although the fitted curves adequately describe the flux variations.

The monochromatic light curves in the 'pseudo-continuum' of 56~Ari
in the near-UV spectral region from 1950 to 3200~\AA\ have a similar
shape: the primary minimum at phase 0.0, a maximum at phase 0.25,
while a secondary minimum and maximum around the phases 0.4-0.5 and
0.6-0.7, respectively. The double-wave light variations in this
spectral region are in antiphase to the variations in the far-UV
spectral region (see Paper~I), but in phase to the variations in the
visual spectral region \citep{Adelman_Fried}. The amplitude of the maximum
at phase 0.25 quickly decreases with increasing wavelength. The second
maximum around the phases 0.6-0.7 increases with increasing
wavelength, as illustrated by Fig.~\ref{differ_flux}.
Thus, the differences of the fluxes at phases 0.25 and 0.0 decrease from 0.5~$\times$~{\units} at $\lambda$~2000~\AA\ to 0.15~$\times$~{\units} at
$\lambda$~3200~\AA. On the other hand, the differences of the fluxes
at phases 0.65 and 0.0 increase only from 0.17~$\times$~{\units} at
$\lambda$~2000~\AA\ to 0.3~$\times$~{\units} at $\lambda$~3200~\AA.
It should be noted, that the standard deviation of the residual
scatter around the fitted curves are equal 0.041 and 0.071 in the units {\units}
on the top and on the bottom panels of Fig.~\ref{differ_flux}, respectively.
Mainly the silicon is responsible for the redistribution of the flux at
phase 0.25 (see Paper~I). On the other hand, in order to explain such behavior
of the monochromatic light curves at phase 0.65 we need to involve other element
than silicon. First of all, we should expect
correlation between the light variations in the 'pseudo-continuum'
and the blocking of the flux by Fe(II) lines, because mainly iron is
responsible for the second minimum of the light curves at phase 0.65
in the far-UV spectral region (see Paper~I).

\begin{figure} [t]
\vspace{1.5mm}
\centerline{\includegraphics[width=83mm]{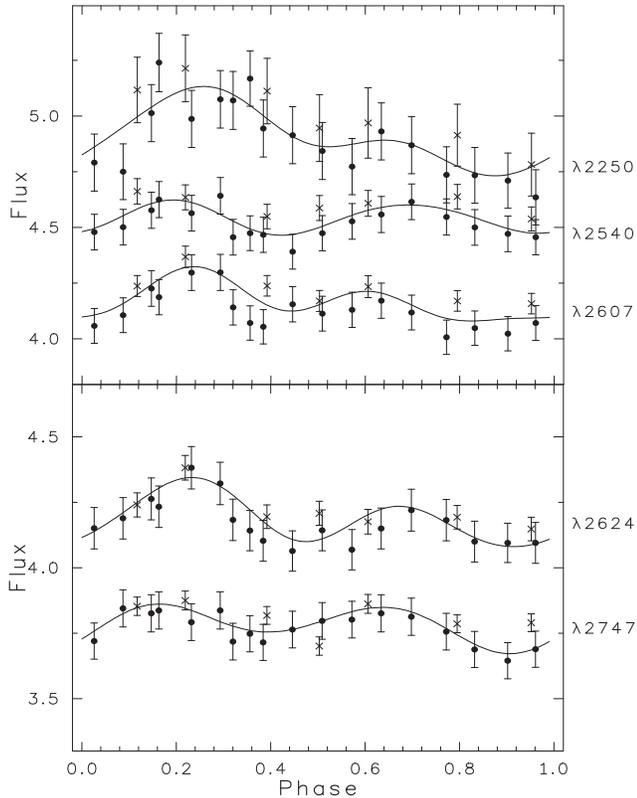}}
\caption{Phase diagrams of the light curves in the
 cores of Fe(II) lines for 56~Ari. Vertical units are presented in {\units}.
 In the {\it all plots}, {\it closed circles} and {\it crosses} represent the data
 from the first and second series, respectively.
 To avoid overlapping the vertical shift on the constant value of the light curve
 at $\lambda$~2540~\AA\ was used (see text).
 The solid lines are the least square fits.
\label{cores_fe1}}
\end{figure}

\begin{figure} [t]
\vspace{1.5mm}
\centerline{\includegraphics[width=75mm]{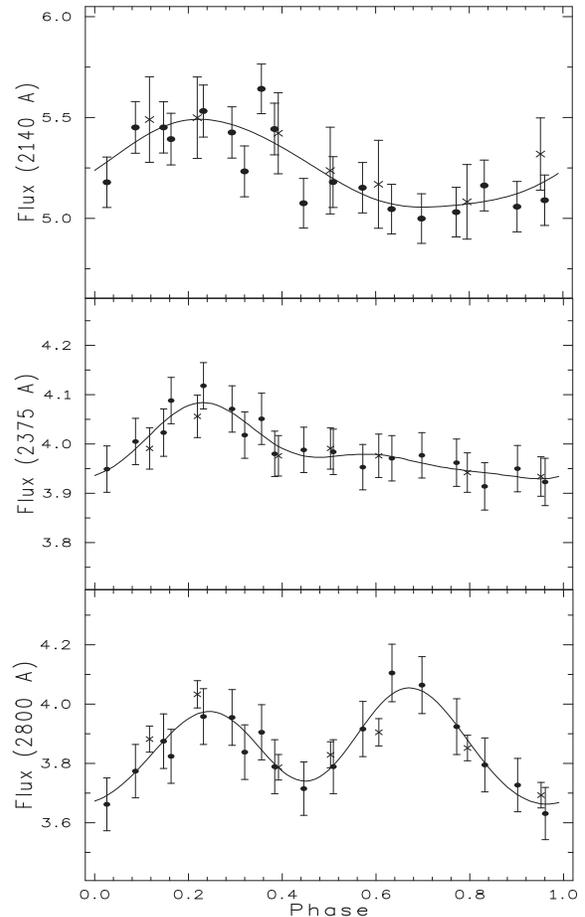}}
\caption{The same as in Fig.~\ref{cores_fe1} for the light curves
 at $\lambda$~2140~\AA, the large feature at $\lambda$~2350-2400~\AA\
 and Mg~{\sc ii} resonance lines.
\label{cores_fe2}}
\end{figure}

\subsection[] {Variation of spectral lines}

The measurement of the equivalent width of a line is one of the best
way to estimate the total amount of energy subtracted from the continuum
by an absorption line.
Unfortunately, the equivalent width depends strongly upon the
knowledge of the intensity of the continuum. According our calculations of
the synthetic spectrum we selected the regions centered at
$\lambda\lambda$~2250, 2540, 2607, 2624 and 2747~\AA\ where Fe(II)
appears as the main absorber. Moreover, we checked stability of the
fluxes in the cores of these regions and found scattering of the
points, especially at phases between 0.5 and 1.0.

In order to measure the absorption in the cores of Fe(II) lines the spectra
were processed using the spectral reduction software {\sc spe} developed
by S. Sergeev at the Crimean Astrophysical Observatory (CrAO).
The program allows measuring the mean intensity of the flux and theirs errors
in any selected rectangular spectral region. In the all cases, these
spectral regions have been chosen $\sim$10~\AA\ wide and centered at
$\lambda\lambda$~2250, 2540, 2607, 2624 and 2747~\AA.
Figure~\ref{cores_fe1} exhibits the variations of the mean intensity and the error
of the flux in the cores of Fe(II) lines versus the rotational phase.
The solid lines are the least squares fits by three-frequency cosine functions.
In order to avoid overlapping, the light curve at $\lambda$~2540~\AA\
was shifted by the value of +0.5~$\times$~{\units}.
It should be noted that the vertical scales differ for each part of the figure.
As can be seen in the graphs of Fig.~\ref{cores_fe1}, all light curves in the cores of Fe(II) lines show the primary minimum at phase 0.0, a maximum at phase 0.25, with a secondary minimum and maximum around the phases 0.5 and
0.65, respectively. The maximal amplitudes of the flux at phases 0.25 and 0.65
show the same behavior as the monochromatic light curves in the 'pseudo-continuum'
of 56~Ari.
Although, the differences of the fluxes at phases 0.65 and 0.0 in
the cores of Fe(II) lines are less than in the near 'pseudo-continuum'
of 56~Ari (see Fig.~\ref{differ_flux}). Probably, the additional blocking
of the flux by many Fe(II) lines occur in these spectral regions.

\section[] {Discussion}

For the first time, \citet{Peterson1970} showed that  the enhanced ultraviolet
opacities produces an opposite variation of the flux distribution in the visual
and the UV spectral regions. Later \citet{Panek1982} noted that flux
redistribution by no single element can explain the light variations of 56~Ari
in the far-UV spectral region.
\citet{Stepien_Czechowski} have investigated the influence of silicon,
helium and iron on the monochromatic light variations in the spectrum of 56~Ari.
The authors noted that more quantitative comparison of the light variations with
the variations of Si(II) $\lambda$4128~\AA\ shows some inconsistency.
They also concluded that the variations of helium seem to have very little
influence on the light variations, and the assumed strictness in phase variation
of iron, as observations suggest, is probably an over-interpretation of the
existing data.

Most probably that the redistribution of the flux from the far-UV to near-UV
spectral regions at phase 0.25 is connected with nonuniform distribution of
silicon on stellar surface of 56~Ari (see Paper I).
On the other hand, the redistribution of the flux at phase 0.65 is quite complex, because there are the additional blocking and redistribution of the flux by iron lines in the near-UV spectral region. Thus, for example, the differences of the fluxes
at phases 0.65 and 0.0 in the cores of Fe(II) lines are less than in the near 'pseudo-continuum' of 56~Ari (see Fig.~\ref{differ_flux}). Moreover, the second maximum at phase 0.65 disappears in the large feature at $\lambda$~2350-2400~\AA\ where a large number of iron lines could explain this feature \citep{Adelman1993}.
Such behavior of the flux in the cores of Fe(II) lines and in the large feature at $\lambda$2350-2400~\AA\ mainly supports that iron is responsible for the redistribution of the flux at phase 0.65. On the other hand, the second maximum is replaced by a minimum in the core of the depression at $\lambda$~2140~\AA,
as illustrated by Fig.~\ref{cores_fe2}. Probably, the additional blocking of the flux by chromium or/and nickel is responsible for such behavior of
the flux at $\lambda$~2140~\AA. It should be noted that according to the VALD-2
database in the UV spectral region the number of the iron lines is significantly
larger than the lines of the other elements.

Probably, an uneven surface distribution of silicon and iron mainly influence on
the flux redistribution in the spectrum of 56~Ari, although an additional sources
of opacity can be involved.
It is not new that silicon and iron influence on the flux redistribution in
the spectra of CP~stars. \citet{Khan_Shulyak2007} have studied the effects of
individual abundance patterns on the model atmospheres of CP stars. They concluded
that the group of elements which produce large changes in the model atmosphere
structure and energy distribution consists of Si, Fe, Cr.
Recently, \citet{Krticka2009} have simulated the light variability of the star HR~7224 using the observed surface distribution of silicon and iron.
They have shown that the silicon bound-free transitions and iron bound-bound transitions provide the main contribution to the flux redistribution, although an additional source of opacity is needed.

It is very interesting to compare our result with the chemical elements distribution
on the surface of 56~Ari obtained in the optical spectral region.
Unfortunately, the only silicon distribution on the surface of 56~Ari was obtained in the optical spectral region by \citet{Ryabchikova2003}. A comparison between
the silicon distribution on the surface obtained from optical spectral region and the results obtained at this paper shows some inconsistency. The Doppler imaging technique
gives a few silicon spots on the surface of 56~Ari
\citep[see Fig.~1 of][]{Ryabchikova2003} while our result is that the silicon, mainly, concentrate at phase 0.25 (Paper~I). Note that the accurate continuum normalization is
crucial for the analysis of the spectral lines in the spectra of CP~stars.
Moreover, the impossibility of fitting the wing and the core of some strong spectral
lines with the same abundance can be observed at the visual spectral region for  CP~stars \citep{Ryabchikova2003}. On the other hand, the light variability can not be easily deduced from the surface distribution without some calculation.

The vertically dependent abundance stratification in chemically peculiar stars are suggested by many authors \citep[e.g.,][and references therein]{Ryabchikova2008}.
This effect may influence our results, because the effective depth at which continuum forms at $\lambda$~2000~\AA\ can differs from  the effective depth at which continuum forms at $\lambda$~3200~\AA.
The vertical abundance stratification itself can also influence the emergent continuum flux. For silicon-rich stars with an effective temperature of about
10000-15000 K, hydrogen becomes the major source of continuum opacity
in the near-UV spectral region, because silicon continuum opacity is
significant in the spectral region with $\lambda~<$~1600~\AA\ \citep{Lanz1996}.
As a result, the maximum amplitude at phase 0.25 is decreasing with increase
wavelength (see Fig.~\ref{differ_flux}).
Probably, the maximum concentration of silicon is in the inner layers of the atmosphere of 56~Ari. On the other hand, behavior of the maximum amplitude at phase 0.65
is quite complex, because there are many iron lines in the near-UV spectral region.
The lines of iron additionally is blocking the flux
at phase 0.65 and, as a result, we see significant scattering of the points on
the bottom panel of Fig.~\ref{differ_flux}. Although, the maximum amplitude
at phase 0.65 is increasing with decrease in wavelength. Probably, the maximum concentration of iron is in the outer layers of the atmosphere of 56~Ari.
It should be noted that the maximum amplitudes at phases 0.25 and 0.65 is showing
the same behavior in the optical spectral region.
For example, the maximum amplitude of the flux at phase 0.65 is in the Balmer
continuum and the maximum amplitude of the flux at phase 0.25 is in the Paschen
continuum for $u$ and $v$ filters of Str{\"o}mgren photometric system, respectively \citep{Adelman_Fried}.

\begin{figure} [t]
\vspace{1.5mm}
\centerline{\includegraphics[width=80mm]{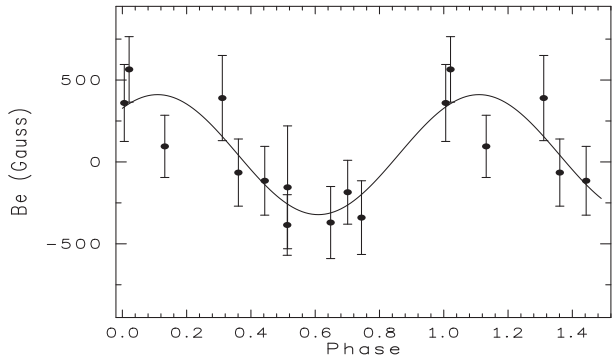}}
\caption{Phase diagram of the effective magnetic field (Be) variations
 for 56~Ari. Magnetic field observations are taken from \citet{Borra1980}.
 The {\it solid line} is the least square fit cosine wave.}
\label{magn_field}
\end{figure}

There is another possibility to explain the flux redistribution from
the UV spectral region to the visual spectral region.
\citet{Kochukhov_et_al}, \citet{Khan_ShulyakA} and \citet{Khan_ShulyakB}
investigated the influence of the polarized radiative transfer and magnetic line
blanketing on the energy distribution, atmospheric structure and photometric
colors. They have shown that the most conspicuous signature of the magnetically
modified line blanketing is a flux deficiency in the UV spectral region and the
respective flux excess in the visual spectral region.
In literature there are the series of the magnetic field measurements obtained by
\citet{Borra1980} for the star 56~Ari. In order to check a possible influence of
the magnetic field on the flux deficiency in the UV spectral region we computed
the phase diagram of the magnetic field variations using
the equation~(\ref{ephemeris}), as illustrated by Fig.~\ref{magn_field}.
One can see that the flux maximum at far-UV (Paper~I) and the flux minimum at
near-UV (Fig.~\ref{mono_light}) occurs about a tenth of a cycle before positive
magnetic extremum. Also, it is impossible to explain the change shape of
the light curves in Balmer and Paschen continua ($u$ and $v$ filters of
Str{\"o}mgren photometric system, respectively) by the magnetic field variation
of 56~Ari. The same result was obtained by \citet{Bohlender1993} at least for
four stars for which extremums of the photometric or spectroscopic variations
do not coincide with magnetic extremums.
\citet{Khan_ShulyakB} have not found any significant changes in model atmosphere
structure, photometric indices, the energy distribution and profiles of hydrogen
Balmer lines that depend on the magnetic field inclination angle.
Note that in the case of 56~Ari the magnetic field is too weak to be
the cause for any considerable light variations \citep{Khan_ShulyakA}.
Additionally, the discovery of mercury spots on the surface of
$\alpha$~Andromedae demonstrate that magnetic field is not the only responsible
for creating and supporting surface structures in CP stars
\citep{Adelman_et_al_2002}.

\section[] {Conclusions}

The archival {\iue} spectrophotometric observations of 56~Ari had permited
to analyze the light variations in the near-UV spectral region from
1950~\AA\ to 3200~\AA\ and to compare them with the variations in
the cores of the spectral features and lines. Although the
double-wave light variations in this spectral region are in
anti-phase to those in the far-UV region, the spectrum of 56~Ari is
variable in the wavelength interval from 1950~\AA\ to
3200~\AA\ over the whole rotational period. The monochromatic light curves
continuously are changing their shape with wavelength.
The exception is the core of the depression at $\lambda$~2140~\AA\
where the second maximum at phase 0.65 is replaced by a minimum and also
the large feature at $\lambda$~2350-2400~\AA\ where the second maximum
disappears. This indicates that we do not observe a truly 'null wavelength region' where the monochromatic light curve has a zero amplitude.
In the Paper~I we concluded that the variations of silicon and iron seem to have influence on the light variations in the far-UV spectral region.
The present investigation mainly supports our previous result, although an
additional sources of opacity can be involved.

\section*{Acknowledgements}
I would like to thank the referee whose criticism and offers which have helped
me the better presentation of the paper.
This research has made by using of $IUE$ Newly Extracted Spectra (INES) data from
the $IUE$ satellite.
%
%\nocite{*}
\bibliographystyle{spr-mp-nameyear-cnd}
%\bibliography{myref}
\bibliography{biblio-u1}

\end{document}